\documentclass[1p]{elsarticle}

\usepackage{graphicx}

\begin{document}

\begin{frontmatter}

\title{Computational diagnosis and risk evaluation for canine lymphoma}

\author{E. M. Mirkes}
\address{Department of Mathematics, University of Leicester, Leicester, LE1 7RH, UK}

\author{I. Alexandrakis, K. Slater, R. Tuli}
\address{Avacta Animal Health, Unit 706, Avenue E, Thorp Arch Estate, Wetherby, LS23 7GA, UK}

\author{A. N. Gorban}
\ead{ag153@le.ac.uk}
\address{Department of Mathematics, University of Leicester, Leicester, LE1 7RH, UK}

\begin{abstract}
The canine lymphoma blood test detects the levels of two biomarkers, the acute phase
proteins (C-Reactive Protein and Haptoglobin).  This test can be used for diagnostics,
for screening, and for remission monitoring as well. We analyze clinical data, test
various machine learning methods and select the best approach to these problems. Three
family of methods, decision trees, kNN (including advanced and adaptive kNN) and
probability density evaluation with radial basis functions, are used for classification
and risk estimation. Several pre-processing approaches were implemented and compared. The
best of them are used to create the diagnostic system. For the differential diagnosis the
best solution gives the sensitivity and specificity of 83.5\% and 77\%, respectively
(using three input features, CRP, Haptoglobin and standard clinical symptom). For the
screening task, the decision tree method provides the best result, with sensitivity and
specificity of 81.4\% and $>$99\%, respectively (using the same input features). If the
clinical symptoms (Lymphadenopathy) are considered as unknown then a decision tree with
CRP and Hapt only provides sensitivity 69\% and specificity 83.5\%.  The lymphoma risk
evaluation problem is formulated and solved. The best models are selected as the system
for computational lymphoma diagnosis and evaluation the  risk of lymphoma as well. These
methods are implemented into a special web-accessed software and are applied to problem
of monitoring dogs with lymphoma after treatment. It detects recurrence of lymphoma up to
two months prior to the appearance of clinical signs. The risk map visualisation provides
a friendly tool for explanatory data analysis.
\end{abstract}
\begin{keyword}
Cancer diagnosis \sep Data analysis \sep Classification \sep Risk evaluation \sep
Decision tree \sep Advanced KNN \sep Radial basis functions
 \PACS 87.10.Vg \sep  87.19.xj
\end{keyword}

\end{frontmatter}

\section{Introduction}\label{introd}

\subsection{Biomarkers for canine lymphoma}

Approximately 20\% of all canine tumours are lymphoma \cite{Vail2001}. The typical age of
a dog with lymphoma is 6-9 years although dogs of any age can be affected. The biggest
problem with cancer treatment in dogs or humans is the earlier diagnostics. Routine
screening can improve cancer care by helping pick up tumours that might otherwise be
missed.

The minimally invasive tests are needed for screening and differential diagnosis as
precursors to histological analysis.  It is also necessary to monitor the late effects of
treatment, to  identify or explain trends and to watch the lymphoma return. The modern
development of veterinary biomarker technology aims to answer these challenges. In the
discovery of cancer biomarkers the veterinary medicine follows human oncology with some
delay. The controversies, potentials biases, and other concern related to the clinical
application of biomarker assays for cancer screening are discussed in \cite{Henry2010}.
There is increasing interest in the study of prognostic and diagnostic biomarker proteins
for canine lymphoma \cite{Mobasheri2013}.

Identification of several biomarkers for canine lymphoma has been reported during the
last decade:
\begin{itemize}
\item{The proteomic evaluation of lymph nodes from dogs with B-cell lymphoma (11
    cases) was compared to those from unaffected controls (13 cases). The expression of
    prolidase (proline dipeptidase), triosephosphate isomerase and glutathione
    S-transferase was decreased in the samples from the lymphoma cases and the
    expression of macrophage capping protein was increased \cite{McCawEtAl2007}.}
\item{The surface-enhanced laser desorption-ionization time-of-flight mass
    spectrometry (SELDI-TOF-MS) was used to identify biomarker proteins for B-cell
    lymphoma in canine serum. 29 dogs with B-cell lymphoma and 87 control dogs were
    involved in the study. Several biomarker protein peaks in canine serum were
    identified, and a classification tree was built on the basis of 3 biomarker
    protein peaks. It was reported that with 10-fold cross-validation of the sample
    set, the best individual serum biomarker peak had 75\% sensitivity and 86\%
    specificity and the classification tree had 97\% sensitivity and 91\% specificity
    for the classification of B-cell lymphoma \cite{GainesEtAl2007}.}
\item{A commercially available canine lymphoma screening test was developed by
    PetScreen Ltd \cite{RatcliffeSlater2009}. Serum samples were collected from 87
    dogs with malignant lymphoma and 92 control cases and subjected to ion exchange
    chromatography and SELDI-TOF-MS analysis. Nineteen serum protein peaks differed
    significantly (p$<$0.05) between the two groups based on normalized ion
    intensities. From these 19 peaks, two differentiating biomarkers emerged with a
    positive predictive value (PPV) of 82\%. These biomarkers were used in a clinical
    study of 96 dogs suspected of having malignant lymphoma. A specificity of 91\%
    and sensitivity of 75\% was determined, with a PPV of 80\% and negative
    predictive value (NPV) of 88\%. Later on, these peaks were identified as two
    acute phase proteins: Haptoglobin (Hapt) and C-Reactive Protein (CRP)
    \cite{Alexandrakis2012}.}
\item{Some qualitative alterations were identified in dogs with lymphoma in the
    proteomic study  \cite{Atherton2013}; 21 dogs included in the study had high
    grade lymphoma confirmed cytologically (16 cases) or histologically (five cases).
    The increased concentrations of haptoglobin in the sera of dogs with lymphoma
    could account for increased levels of $\alpha$2 globulins, $\alpha$2
    macroglobulin, $\alpha$-anti-chymotrypsin and inter-$\alpha$-trypsin inhibitor,
    which were identified concurrently.}
\item{Vascular endothelial growth factor (VEGF), metalloproteinase (MMP) 2 and 9
    transforming growth factor beta (TGF-$\beta$)  were tested  in 37 dogs with
    lymphoma, 13 of which were also monitored during chemotherapy. Ten healthy dogs
    served as control. Lymphoma dogs showed higher activity of MMP-9 (p$<$0.01) and
    VEGF (p$<$0.05), and lower TGF-$\beta$ than controls, and a positive correlation
    between act-MMP-9 and VEGF (p$<$0.001). During chemotherapy, activity MMP-9 and
    VEGF decreased in B-cell lymphomas (p$<$0.01), suggesting a possible predictive
    role in this group of dogs \cite{AresuEtAl2014}.}
\end{itemize}

For use in clinics, the biomarkers should be identified and validated in preclinical
settings and then validated and standardized using real clinical samples
\cite{Mobasheri2010}. Intensive search of biomarkers requires standardisation of this
technology \cite{Matharoo-Ball2008}. Proteins discovered in the research phase may not
necessarily be the best diagnostic or therapeutic biomarkers. Therefore, after
identification of a biomarker (Phase 1), the clinical assays are necessary to investigate
if the biomarker can truly distinguish between disease versus control subjects (Phase 2).
Then  special retrospective and prospective research is needed for sensitivity and
specificity analysis (Phases 3 and 4). Finally, the cancer control phase is needed (Phase
5) to ``evaluate role of biomarker for screening and detection of cancer in large
population'' \cite{Matharoo-Ball2008}. Discovery and identification of a promising
biomarker does not mean that it will successfully go through the whole standardised
procedure of testing and evaluation.

\subsection{Acute phase proteins as lymphoma biomarkers}

Acute phase proteins are now understood to be an integral part of the acute phase
response which is the cornerstone of innate immunity \cite{Cray2012}. They have been
shown to be valuable biomarkers as increases can occur with inflammation, infection,
neoplasia, stress, and trauma. All animals have acute phase proteins, but the major
proteins of this type differs by species. Acute phase proteins have been well documented
in laboratory, companion, and large animals. After standardized  assays, these biomarkers
are available for use in all fields of veterinary medicine as well as basic and clinical
research \cite{Cray2012}.

Acute phase proteins including alpha 1-acid glycoprotein
\cite{Ogilvie1993,Hahn1999,Tecles2005}, C-Reactive Protein (CRP)
\cite{Merlo2007,Mischke2007,RatcliffeSlater2009,Alexandrakis2012}, and Haptoglobin (Hapt)
\cite{Mischke2007,RatcliffeSlater2009,Alexandrakis2012}, have been evaluated as tumor
markers. Nevertheless, as is mentioned in review \cite{Henry2010}, it is still
necessary to prove that these biomarkers are clinically useful in cancer diagnosis. Some
authors even suggest that the non-specific serum  biomarkers indicate inflammatory
response rather than cancer \cite{Jesneck}.

In our research we evaluate the role of two biomarkers, CRP and Hapt, for screening and
detection of lymphoma, for differential diagnosis of lymphoma and for monitoring of
lymphoma return after treatment. Our research is based on the PetScreen Canine Lymphoma
Blood Test (cLBT). This is advanced technology to detect lymphoma biomarkers present in a
patient's serum \cite{Alexandrakis2012}. The cLBT evaluates the concentration of two
acute phase proteins: Hapt and CRP. High levels of these biomarkers indicates a high
likelihood that the patient has lymphoma. The cLBT provides a minimally invasive
alternative to a fine needle aspirate as a precursor to histological diagnosis of the
disease. The cLBT should be used for differential diagnosis when a patient is suspected
of having lymphoma by showing classical symptoms such as generalized lymphadenopathy,
PU/PD and lethargy (we call all such cases the {\em clinically suspected} ones). It may
be also useful in the monitoring of lymphoma return. In summary, the test provides:

\begin{itemize}
\item A simple blood test requiring only 2ml of blood taken as part of existing
    biochemistry/haematology work up. Results are available the same day.
\item A minimally invasive procedure.
\item An alternative to taking an FNA sample and the associated risks of failing to
retrieve sufficient lymphoid cells or encountering poor preservation of the cells.
\item A monitoring tool to assess treatment progression and to detect recurrence.
\end{itemize}

Some of our previous results of canine lymphoma diagnosis are announced in
\cite{Alexandrakis2012,MirkesAl2014}.

\subsection{The structure of the paper}

The description of the database and statement of the problems are represented in
Section~\ref{database}.  Two cohorts are isolated in the database and two problems are
formulated: (i) differential diagnostic in clinically suspected cases and (ii) screening.
The isolation of the clinically suspected cohort is necessary for formulation of the
problem of differential diagnostics and selection of the appropriate methods. The healthy
cohort and formulation of the screening problem demands the use of a prior probability of
lymphoma and forbids the use of class weights as a parameter to select the best solution.
This means that the weights of classes are determined by the prior probability. Both
problems (differential diagnostics and screening) are formulated as problems of
probabilistic risk evaluation \cite{BedfordCooke2001}. Usual classifiers provide a
decision rule and give the answer in the form ``Yes'' or ``No'' (cancer or not cancer,
for example). We almost never can be sure that this ``Yes'' or ``No'' answer is correct.
Therefore the evaluation of probability may be more useful than just a binary answer. If
we evaluate the posterior probability of lymphoma under given values of features then we
can take the decision about the next step of  medical investigation or treatment.
Probabilistic risk evaluation supports decision making and allows to evaluate the
consequences of the decisions (risk management \cite{BedfordCooke2001}).

Section \ref{methods} presents  a brief review of the data mining methods employed in biomarker cancer diagnosis.
We introduce the methods used in our work for the analysis of canine lymphoma. The detailed description of these methods
is given in Appendix.
Three used methods are described:
\begin{itemize}
\item {\it Decision trees} with three different impure-based criteria: information gain, Gini
gain and DKM \cite{RokachMaimon2010}.
\item \emph{K nearest neighbors} method (KNN). Three versions of KNN methods are used: KNN with Euclidean distance
\cite{Clarkson2005}, KNN with Fisher's distance transformation, and the advanced adaptive
KNN \cite{HastieTib1996}. All the three methods use statistical kernels to weight an
influence of each of the k nearest neighbors to evaluate the risk of lymphoma. The KNN
method with Fisher's distance transformation is much less known. We use the geometrical
complexity \cite{Zinovyev2013} for comparison of different KNN methods.
\item \emph{Probability density function estimation} (PDFE) \cite{Scott1992}. We use
PDFE for direct evaluation of the lymphoma risk.
\end{itemize}

The decision trees and KNN classifiers are also used for evaluation of probability. The way back from the probability
estimate to classification rule is simple, just define the threshold. The criterion of selection of the best
classifier is the maximum sum of sensitivity and specificity or the furthest from the
``completely random guess'' classifier. We also compare performance of this selection criterion with some other criteria:
the relative information gain (RIG) from the classifier output to the target attribute,  accuracy,
precision, and $F$-score.

We use classical methods, and the main building blocks of the algorithms are well known. Nevertheless, some particular
combinations of methods may be new, for example, combination of discriminant analysis with Advanced KNN (see Appendix).
We have tested automatically thousands of combinations, and the best combination for each task has been selected.

Section~\ref{results} contains the description of the best solutions obtained for
differential diagnostic and screening problems. All features are analyzed from the point
of view of their usability for the lymphoma diagnostic and risk evaluation.  We present
the case study for both problems: for the diagnostics problem we have tested 25,600,000
variants of the KNN method, 5,184,400 variants of decision tree algorithms and 3,480
variants of the PDFE method; for the screening task we have tested 51,200 variants of KNN
and advanced KNN parameters, 10,368 variants of decision trees and 3,480 variants of
PDFE. The versions differ by impurity criteria, kernel functions, number of nearest
neighbors, weights and other parameters. The best results are implemented in web-accessed
software for the diagnosis of canine lymphoma (implemented in Java 6).

The obtained results provide the creation of a more reliable diagnostic, screening and
monitoring system for canine lymphoma. The first application of the developed system shows
that the risk of lymphoma (cLBT score) defined after lymphoma treatment allows prediction
of time before relapse of lymphoma. If after treatment of lymphoma the cLBT is performed
regularly, it detects recurrence up to two months prior to the appearance of physical
signs.

\section{Database description and problem statement}\label{database}

\subsection{Database}

The original database contains 303 records (dogs) with four categorical input features:
Sex, Lymphadenopathy, Neutered and Breed and three real valued features: Age and
concentrations of two acute phase proteins: Haptoglobin (Hapt) and C-Reactive Protein
(CRP). A part of serum samples was collected by PetScreen from dogs undergoing
differential diagnosis for lymphoma and also collected at veterinary practices in the USA
\cite{Alexandrakis2012,RatcliffeSlater2009}. Another source  is the Pet Blood Bank which
stores the blood of healthy dogs. Lymphoma positive serum samples were confirmed either
by excisional biopsy or fine needle aspirate and non-lymphoma serum samples were
confirmed to be free of lymphoma at a minimum of 6 months after the sample was taken
\cite{Alexandrakis2012,RatcliffeSlater2009}.

Breed may be important for lymphoma diagnosis. For example, the boxer, bulldog and bull
mastiff breeds have a high incidence of lymphoma \cite{Pastor2009}. The relatively small
number of records in our database has limited our ability to detect breeds with an
elevated risk. We exclude this feature because there are 54 different breeds in 204
records (less than four records of each breed) and 99 missed values. This amount of known
data  for a categorical feature with 54 different values is not sufficient for diagnosis
without clustering of breeds (numerosity reduction is needed). The well-developed
imputation methods \cite{Saar-Tsechansky} also cannot be applied directly without
numerosity reduction because of insufficient information.

The target feature Lymphoma is binary: ``Positive'' for a dog with lymphoma and ``Negative'' for a dog
without lymphoma. Three attributes contain missed values: Sex contains 96 (35\%);
Neutered contains 107 (38\%); Age contains 101 (36\%).

\subsection{Two cohorts and two problems}

\textit{Isolating of two cohorts.} The database analysis shows that the samples are
heterogeneous: two different cohorts of data can be distinguished in the database. There
were two different sources of data: dogs undergoing differential diagnosis for lymphoma
and the Pet Blood Bank (the blood of healthy dogs) \cite{Alexandrakis2012}.

The existence of two so different sources of data entails the presence of two different
cohorts of patients in the database. The first cohort is entitled ``clinically suspected'' and
contains records collected by PetScreen from dogs undergoing differential diagnosis. All
dogs in this cohort have been referred for differential diagnosis by veterinary
practitioners. The vets decide that these dogs are clinically suspected on the base of
one or more clinical symptoms. It is not possible to find a posteriori these symptoms for
each instance and we have to introduce a new synthetic attribute: ``clinically
suspected''. The cohort of clinically suspected instances should be considered separately
for differential diagnosis purposes and we propose to treat each case referred to the
differential diagnosis as a clinically suspected one. The second cohort is entitled
``healthy'' and contains records obtained from healthy dogs courtesy of the Pet Blood
Bank.

The additional confirmation of existence of two cohorts
is the differences in statistics of the attributes for these cohorts.
In accordance with expert estimations, the prior probability of lymphoma
is located between 2\% and 5\% in the canine population. The number of records of
patients with lymphoma is 97 or 32\% of all the records in the database. All
these cases have been clinically suspected and form 42\% of the clinically suspected
cases. This imbalance entails the usage of specific methods to solve screening tasks.
The ``clinically suspected'' feature was added to the database to identify the two
cohorts. The values of feature ``clinically suspected'' were defined by using additional
information from veterinary cards.

The existence of the two cohorts allows us to formulate two different problems:
the problem of differential diagnosis and the problem of screening.

\textit{Differential diagnosis.}
The problem of differential diagnostic can be formulated as a problem of lymphoma
diagnosis for patients with some clinical symptoms of lymphoma. To solve this task we use
the clinically suspected samples. A diagnostic problem is a usual classification problem
and all classification methods can be used. We use three types of classification methods:
KNN, decision tree and the method based on probability distribution function estimation.
Each of these methods is described in Section~\ref{methods}. The first two
methods have an auxiliary parameter ``weight'' of the positive class $w_{\rm p}$.

\textit{Screening.}
The problem of screening can be formulated as a problem of evaluation of lymphoma risk for
any dog. The sample for this problem includes all the database records. The experts'
estimation of prior probability of lymphoma is between 2\% and 5\% however the fraction
of patients with lymphoma records in the database is 32\%. To compensate for this
imbalance all methods take into account the prior probability of lymphoma and
the weights of classes are defined by prior probability.

\section{Methods}\label{methods}

\subsection{Data mining methods for biomarker cancer diagnosis \label{methodsReview}}

Extraction diagnostic biomarkers for cancer, their validation and testing for clinical use is considered now as a data
analysis challenge \cite{Hilario2004}. The classical methods of supervised classification
are widely used to meet this challenge: linear and quadratic discriminant analysis
\cite{Baggerly,ComSysBioCancer2012,Lilien2003,Wagner2003}, decision trees
\cite{Adam2002,Becker2004,Hilario2003,Markey2003,Neville2003,Prados2004,SuShen2007,Thomas2006,YuChen2005},
logistic regression \cite{Asiago2010,ComSysBioCancer2012,Koopmann2004,Kozak2003}, $k$ nearest neighbors (KNN)
approach \cite{Hilario2003,Prados2004,Wagner2003,Wagner2004} and na\"{\i}ve Bayes model
for probability density function estimation \cite{ComSysBioCancer2012,Ostroff2010}. Artificial neural
networks are used for the identification of cancer biomarkers and cancer prediction  as a
flexible tool for supervised learning
\cite{Ball2002,Hilario2003,Lancashire2005,Prados2004,Tatay2003,Thomas2006,Lancashire2009}.
During the last decade, applications of support vector machines
\cite{Guan2009,LiTang2004,Prados2004,Wagner2003}, and ensemble learning (random forests,
committees of decision trees, boosting methods)
\cite{Cima2011,LiLiu2003,LiTang2004,NamChung2009,WuAbbot2003,Yasui2003} have been
intensively developed.

Most of the works combine and compare several methods, for example, discriminant
analysis, KNN and support vector machines \cite{Wagner2003}, decision trees, KNN, and
artificial neural networks \cite{Hilario2003}, discriminate analysis, random forest, and
support vector machine \cite{NamChung2009}, decision trees, bagging, random forests,
extra trees, boosting, KNN, and support vector machines \cite{Geurts2005}, linear
discriminant analysis, quadratic discriminant analysis, KNN, bagging, boosting
classification trees, and random forest \cite{WuAbbot2003}.

Supervised classification and regression methods are combined with dimensionality
reduction methods such as linear and non-linear principal component analysis
\cite{ComSysBioCancer2012,GorbanKegl2008,Hilario2008,Kirmiz2007,Thomas2006} or moment-based approach
\cite{Tang2010}. Several hybrid systems are developed with combinations of supervised
classification and unsupervised clustering \cite{ComSysBioCancer2012,PYang2010}.

The classical decision trees or KNN approach (or both) usually serve as bases for
comparison when evaluating supervising classification. It is necessary to stress that
there are many versions of algorithm even for a single decision tree or KNN. In this
paper, we systematically test many versions of these basic algorithms on the problem of
canine lymphoma differential diagnosis and screening.

We use three types of classification methods to evaluate the risk of lymphoma for the
problems of differential diagnosis and screening: decision tree, KNN and PDFE. Each of
these titles covers many different algorithm. Detailed description of these families of
algorithms used is presented in Appendix. We aim to select the best one for the given
problem. Simultaneously the best subset of input attributes should be selected.

Totally we have tested 10,368 trees for the screening problem. For the task of
differential diagnostic we vary the weight of class of patients with lymphoma from 0.1 to
50. For the differential diagnostic problem 5,184,400 variants of decision trees have
been tested.

We have tested 51,200 sets of parameter values for the screening. For the differential
diagnostic, we vary the weight of class of patients with lymphoma from 0.1 to 50;
25,600,000 variants of KNN method have been tested.

We have tested 3,840 variants of PDFE for each problem.

\subsection{Data transformation, evaluation and weighting}

The CRP and Hapt features are the concentrations of the two proteins. It is well-known
that in many chemical applications the logarithm of concentration (the chemical
potential) is more informative and useful then the concentration itself
\cite{Westerhoff2005}. Therefore, we test all the methods for concentrations of CRP and
Hapt (in the ``natural'' units of concentration) and for logarithms of the concentrations
(in the logarithmic coordinates). All real valued features are divided by their standard
deviation. If CRP and Hapt are used in logarithmic transformed form then initially we
perform logarithmic transformation and then divide by the standard deviation of the
logarithmic transformed feature. For the KNN and PDFE all the binary input features are
coded by 0 and 1.

For feature evaluation and selection we calculate the \emph{Relative Information Gain}
(RIG) \cite{RokachMaimon2010} which is the natural tool
to estimate the importance of input features for the categorical target feature. For this
purpose, real data have been binned (organized into groups).

We use two types of {\em weights}: prior weights of classes and weight of positive class.
Really, we use the weights of instances instead of weights of classes. For the
differential diagnosis and screening problems both types of weights are defined for
different reasons: for the screening problem we have the prior probability of lymphoma
for the whole dog population; for the differential diagnosis problem we have no prior
probability but can use the auxiliary weight of the positive class to search for the best
classifier. We use the following notations: $p$ is the prior probability of lymphoma,
$N_{\rm L}$ is the number of patients with lymphoma, $N_{\rm CS}$ is the number of all
clinically suspected patients and $N_{\rm H}$ is the number of healthy patient.

For the screening problem the weight of the class of patients with lymphoma is equal to
$p$. The weight of one patient with lymphoma is equal to $w_{\rm L}=p/N_{\rm L}$. In
fact, this is the weight of any record of the clinically suspected cohort. The total
probability must be equal to 1. This means that the sum of weights of all records must be
equal to 1. Therefore, the weight of each record of a healthy patient is $w_{\rm
H}=(1-w_{\rm L} N_{\rm CS})/N_{\rm H}$. For the screening problem the auxiliary weight of
the positive class cannot be used (is equal to 1).

For the differential diagnosis problem there is no prior probability.
The auxiliary weight of positive class may be any positive number.

To work with imbalanced dataset we employ two {\em data simulation methods} for over-sampling of the minority class.

The first approach ({\em ``Rectangular''}) uses the random generation of $N$ new instances for each given sample from the minority class by formulas
\begin{equation}\label{Rectangular}
x_{\rm new}=x + \sigma_x W r_x
\end{equation}
where $\sigma_x$ is the standard deviations, $r_x$ is a random variable uniformly
distributed in interval (-1,1), $W$ is the average Euclidean distance from the given
sample to $k$ nearest neighbors of the same class. (The Euclidean distance is calculated
in the plane of dimensionless variables normalized to unite variance.)

The second approach is synthetic minority over-sampling technique ({\em SMOT})
\cite{Chawla2002}. It also uses the random generation of $N$ new instances for each given
sample from the minority class. For a given $k$, we find $k$ nearest neighbors of the
given sample of the same class. Each new instance is randomly situated on the straight
line interval which links the given sample with a randomly selected nearest neighbor
(from $k$ neighbors found).

\subsection{Selection of the best algorithms}

We have many algorithms (variants of algorithm parameters) and we need to select the best
algorithm. In this study we have considered two possible approaches: (i) use of the test
set and (ii) {\em Leave One Out Cross Validation} (LOOCV). The preference for using test
set is the speed: for each algorithm one model construction is sufficient. The model
construction means the forming of the decision tree, or identifying  $k$ nearest
neighbors, and computing the inverse covariance matrix for PDFE. LOOCV is more expensive:
the number of model constructions is equal to the number of instances. Nevertheless, for
a relatively small sample exclusion of a sufficiently large test set from learning may
lead to the strong scattering of the evaluation result. We split the database into
training set (80\%) and test set (20\%) 100 times independently and find large variance
of the estimated sensitivity and specificity. The values vary  from 30\% to 100\%, and
the best version of the algorithm cannot be defined unambiguously. Therefore, we use
LOOCV for evaluation of sensitivity and specificity and for selection of the best
algorithm. An extensive simulation study of cross-validation for three classification
rules, Fisher's discriminant analysis, 3NN, and decision trees (CART) using both
synthetic and real breast cancer patient data was performed in \cite{Braga-Neto2004}. It
was demonstrated that cross-validation is less biased than some other methods but
overestimates the number of errors for small samples.

The next question is what indicator has to be used as a measure of algorithm accuracy. We
can calculate the accuracy of classification as a ratio of correctly classified cases
among all cases. Also the sensitivity and specificity can be used as such a measure. The
classification accuracy is appropriate when numbers of examples of different classes are
balanced. In our database the fraction of lymphoma patients is equal to 32\% for the
screening problem and 42\% for the differential diagnosis problem. This means that the
algorithm selected by classification accuracy can be shifted to good specificity and poor
sensitivity. Other commonly used measures of classification quality is the area under
Receiver Operating Characteristic (ROC) \cite{Tanner1954}. The sum of specificity and
sensitivity is the distance from curve to the main diagonal which corresponds to the
`completely random guess' classifier. We suggest considering the classifier with maximum
sum of specificity and sensitivity as the best.

\section{Results}\label{results}

\subsection{Feature evaluation by information gain}

We need to find how much information about the diagnosis contain the inputs. For this
purpose, real data have been binned (organized into groups). The bins have approximately
equal depth; the boundaries of bins are represented in Table \ref{tab1}. Table
\ref{tab2} contains values of RIG for the target feature Lymphoma from any input feature.
RIG is calculated for the whole database and for two samples: (Y) with L=``Y'' and (N)
with L=``N'' (we use the abbreviation L for Lymphadenopathy). The RIG from Neutered is
always less than 1\% and this feature is excluded from the further study.

We calculate the RIG for Lymphoma from all the input features together. The calculated
value of RIG is 83\%. This gives us an estimate of the expected classification accuracy.
Therefore, we do not expect to produce classifiers without misclassifications.

\begin{table}{\footnotesize\begin{center}
\caption{\label{tab1} Real attributes and bins.}
\begin{tabular}{cccc}
\hline
Feature & min &max & Upper bounds of bins  \\   \hline
Age & 0.67 & 17 & 3, 6, 8, 11, 20 \\  \hline
CRP & 0 & 124 & 0.6, 2.5, 11, 27, 125 \\   \hline
Hapt & 0& 18 & 0.2, 1.7, 4, 7.5, 20 \\    \hline
\end{tabular}
\end{center}}
\end{table}

\begin{table}{\footnotesize
\begin{center} \caption{\label{tab2}Relative information gain about the ``Lymphoma'' feature.}
\begin{tabular}{cccc}\hline
& & \multicolumn{2}{c}{RIG under given L}\\ \hline
Tested feature&RIG& L=Y& L=N\\  \hline
 L&28.92\%& -- & --\\	 \hline	
CRP binned&24.38\%&15.00\%&23.52\%\\  \hline
Hapt binned&\ 07.02\%&\ 01.76\%&14.32\%\\   \hline
Age binned&\ 06.07\%&\ 01.62\%&\ 09.39\%\\   \hline
Sex&\ 00.95\%&\ 03.79\%&22.84\%\\            \hline
Neutered&\ 00.06\%&\ 00.50\%&\ 00.47\%\\     \hline
\end{tabular}
\end{center}}
\end{table}

The number of input attributes is five and the problem of feature selection can be solved by exhaustive
search. Table \ref{tab2} shows that the most informative attributes are Hapt (H) and
CRP (C). These features are included into all tested sets of input features. The various
combination of Age (A), Sex (S) and Lymphadenopathy (L) are included into tested input
sets. In total, eight input feature sets are formed. Each set is denoted by abbreviation
of included features: CH, CHA, CHL, CHS, CHAL, CHAS, CHLS and CHALS.

\begin{figure}[t]
\centerline{
\includegraphics{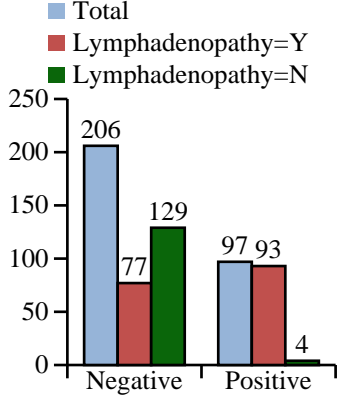}}
\caption{\label{fig1} Distribution of Lymphoma diagnosis.}
\end{figure}

The distributions of the real valued features are non-normal. It means that we cannot use
any methods based on assumption of normality. The distribution diagram for Lymphoma is
represented in Fig.~\ref{fig1}. The diagram shows that only four records without
lymphadenopathy have a positive Lymphoma diagnosis. It means that the decision ``all dogs
without lymphadenopathy have no lymphoma'' generates only 4 false negative errors and
truly identifies 93 dogs without lymphoma.

\subsection{The best algorithms}

The criteria developed for choosing the best solution suggest selecting the following
algorithms. The ROC curves for the selected classifiers are depicted in Fig.\ref{fig2}.

\textit{Differential diagnostic problem.} The best algorithm is the decision tree with
three input features: a linear combination of the concentrations of CRP and Hapt, and
Lymphadenopathy. The tree is formed with DKM as the splitting criterion. The sensitivity
of this method is 83.5\%, specificity is 77\%. The ROC integral for this method is 0.879
(Fig.~\ref{fig2}a).

In the case when Lymphadenopathy is considered as unknown we use a decision tree which
only uses CRP and Hapt. The tree is formed with Information gain as a splitting
criterion. The best version uses input features in linear combinations after logarithmic
transformation. The sensitivity of this method is 81.5\%, the specificity is 76\%. The
ROC integral (Fig.~\ref{fig2}b) for this method is 0.780.

\textit{Screening.} The best classifier for the screening problem is the decision tree
with three input features: the concentrations of CRP and Hapt, and Lymphadenopathy. The
tree is formed with DKM as a splitting criterion. The concentrations of CRP and Hapt are
used separately (not in linear combinations). The sensitivity of this method is  81.4\%
and  specificity is $>$99\% (no false negative results in one-leave-out
cross-validation). The ROC integral is 0.917 (Fig.~\ref{fig2}c). In the case when
Lymphadenopathy is considered as unknown we use a decision tree with CRP and Hapt only.
The tree is formed with Gini gain as the splitting criteria. The concentrations of CRP
and Hapt are used separately. The sensitivity is 69\%, the specificity is 83.5\%. The ROC
integral is 0.771 (Fig.~\ref{fig2}d).

\begin{figure}[t]
(a)\includegraphics[width=.18\textwidth]{figure2a.eps}$\;\;$
(b)\includegraphics[width=.18\textwidth]{figure2b.eps}$\;\;$
(c)\includegraphics[width=.18\textwidth]{figure2c.eps}$\;\;$
(d)\includegraphics[width=.18\textwidth]{figure2d.eps}
\caption{\label{fig2}ROC curves for (a) the best algorithm for differential diagnosis
(ROC integral 0.879), (b) the best algorithm for differential diagnosis with CRP and Hapt
only (ROC integral 0.780), (c) the best algorithm for screening (ROC integral 0.917), and
(d) the best algorithm for screening with CRP and Hapt only (ROC integral 0.771).}
\end{figure}

The classifiers for screening are prepared using the mixture of clinically suspected patients (Lymphadenopathy=Y)
with the patients without lymphadenopathy. Application of these classifiers for screening of dogs without clinical symptoms (Lymphadenopathy=N)
requires additional tests because there are only four cases of dogs with lymphoma with Lymphadenopathy=N in the database (see Fig.~\ref{fig1}).
For the preliminary analysis of this problem we apply both data simulation methods for over-sampling of the minority class, Rectangular (\ref{Rectangular}) and SMOT.

We select $N=10$ and $k=3$ in each method and add synthetic data to the instances from
the original database with Lymphadenopathy=N and positive lymphoma diagnosis. The new
database has well balanced classes. For both methods, the two best decision trees (one for
Lymphadenopathy=Y and one for Lymphadenopathy=N) together demonstrate in LOOCV the
results presented in Table~\ref{tab3}. We see that with the simulated data
specificity decreases. Therefore, it is desirable to collect more instances with lymphoma
but without observable lymphadenopathy (Lymphadenopathy=N)  for the validation of
screening algorithms.

\begin{table}{\footnotesize\begin{center}
\caption{\label{tab3} Sensitivity and specificity for extended dataset.}
\begin{tabular}{ccc}
\hline
Method & Sensitivity & Specificity   \\   \hline
Rectangular & 89.1 & 62.2  \\  \hline
SMOTE & 88.3 & 65.2 \\   \hline
\end{tabular}
\end{center}}
\end{table}

\textit{Other criteria.}
The value of the Hosmer-Lemeshow \cite{Hosmer2000} statistics for the differential
diagnosis algorithm with three input value (CHL) is 12.73. It shows that with $p$-value
greater than 10\% the distribution of estimated probabilities coincides with the
distribution of diagnosis. This test does not consider the prior probability and cannot
be applied for the screening problem. Efron's pseudo $R^2$ \cite{Efron1978} shows that
classifiers which use Lymphadenopathy explain about 40\% of total variance. McFaden's
pseudo $R^2$ \cite{McFadden1974} for the differential diagnosis problem classifier, which
uses Lymphadenopathy, has 38\% greater log likelihood than the null model ones. For the
screening problem the classifier which uses Lymphadenopathy has 45\% greater log
likelihood than the log likelihood of null model which is based on prior probability.

We employ the {\em Sensitivity + Specificity} criterion for the best model selection.
There exist many other criteria, for example, {\em relative information gain} (RIG) from
the classifier output to the target attribute (Lymphoma, in our case), {\em Accuracy}
([``True positive'' + ``True negative'']/"Number of instances"), {\em Precision} (``True
positive''/``Number of positive labels''), where ``Number of positive labels'' is the
number of samples labeled as positive, i.e. ``True positive'' + ``False positive'', {\em
$F$-score} that is the harmonic mean of Precision and Sensitivity ($F$=
2$\times$Precision$\times$Sensitivity/[Precision + Sensitivity]). We compare performance
of these criteria on the test task of selection of the best model for the data set CHL
without logarithmic transformation of concentrations. Table~\ref{tab4} represents the
sensitivity (Sens) and specificity (Spec) for the best models which are selected by each
criterion.

\begin{table}{\footnotesize\begin{center}
\caption{\label{tab4} Sensitivity and specificity (\%) for the best models selected by
different criteria}
\begin{tabular}{ccccccccccc}
\hline
 & \multicolumn{2}{c}{Sens+Spec} & \multicolumn{2}{c}{RIG} & \multicolumn{2}{c}{Accuracy} & \multicolumn{2}{c}{Precision} & \multicolumn{2}{c}{$F$-score} \\ \hline
 Method & Sens & Spec & Sens & Spec & Sens & Spec & Sens & Spec & Sens & Spec \\  \hline
 DT & 83.5 & 77.0 & 83.5 & 77.0 & 79.4 & 79.3 & 78.4 & 80.0	& 83.5 &77.0 \\   \hline
 KNN & 79.4 & 75.6 & 84.5 & 70.4 & 79.4 & 75.6 & 4.1 & 100.0 & 84.5 & 70.4 \\    \hline
 PDFE & 83.5 & 68.9 & 83.5 & 68.9 & 77.3 & 74.8 & 70.1 & 78.5 & 83.5 & 68.9   \\    \hline
\end{tabular}
\end{center}}
\end{table}

As we can see from this test, only the criterion Precision sometimes gives
significantly different results (very precisely all the positive labels are true positive
but many false negative results occur). All other criteria produce similar results.

\textit{Risk evaluation and risk mapping.} All classifiers used in our study can calculate the risk of lymphoma at an arbitrary
point. We can use this capability to form a map of risk. To visualize data with more than
two dimensions several types of screens can be used: coordinate planes, PCA, non-linear
principal graphs and manifolds \cite{GorbanKegl2008,GorbanZin2010}. For this study we use
the plane of CRP and Hapt concentrations. The explanation of colours is depicted in the
legend included at the right of each figure.

\begin{figure}[t]
(a)\includegraphics[width=.25\textwidth]{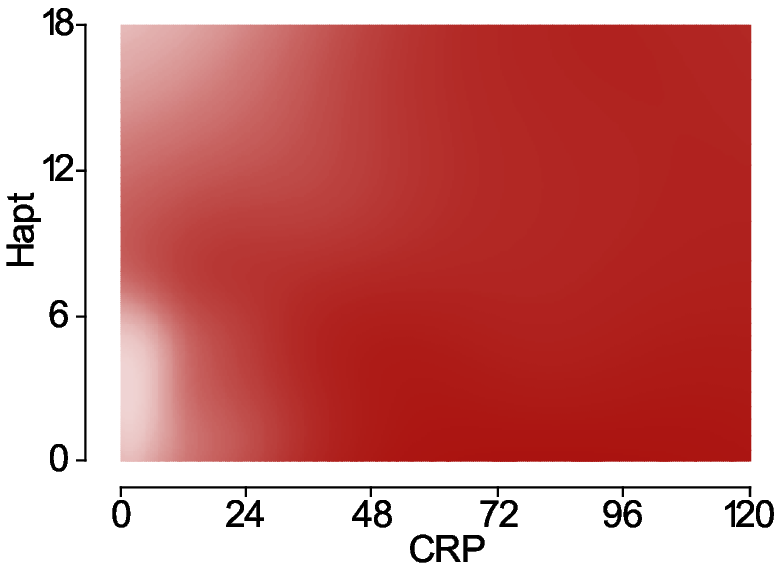}
(b)\includegraphics[width=.25\textwidth]{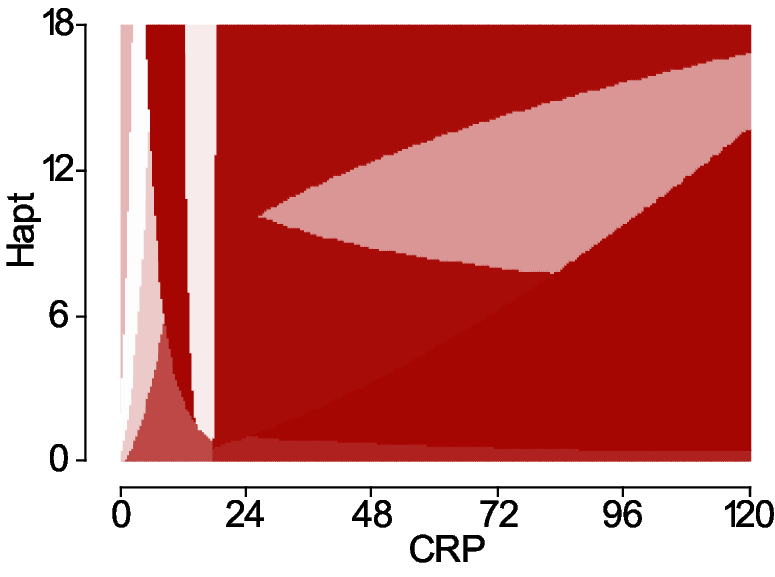}
(c)\includegraphics[width=.25\textwidth]{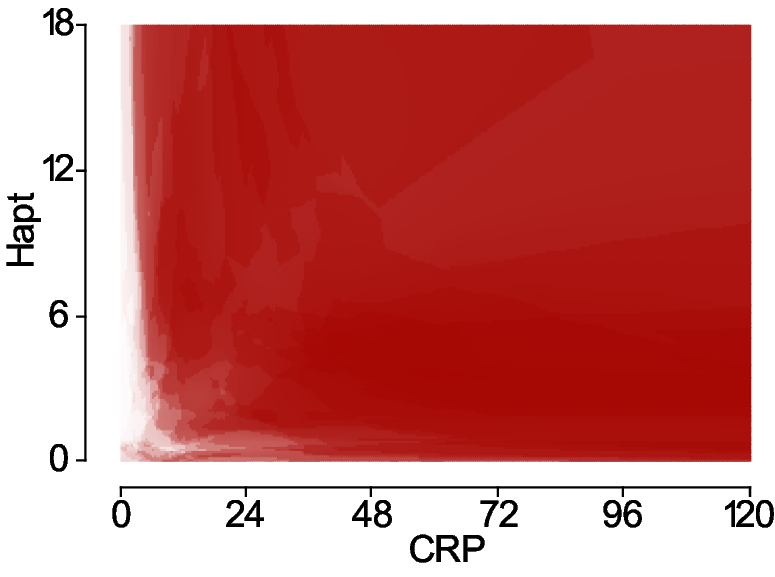}
(d)\includegraphics[width=.048\textwidth]{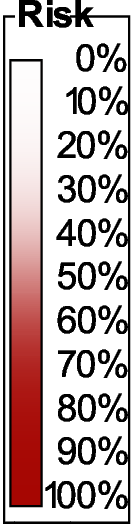} \\
(e)\includegraphics[width=.25\textwidth]{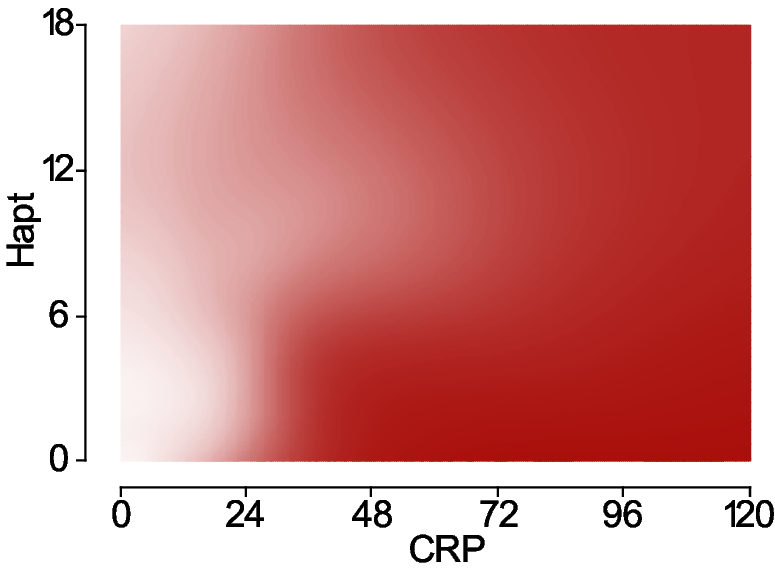}
(f)\includegraphics[width=.25\textwidth]{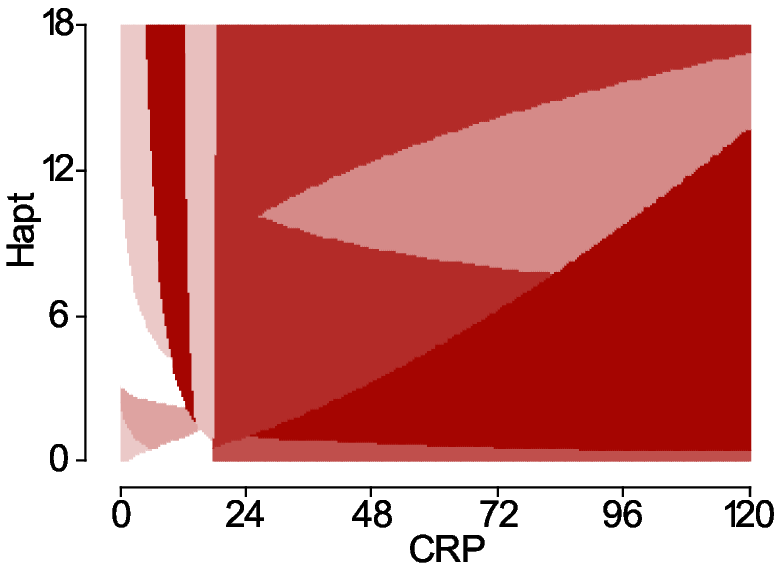}
(g)\includegraphics[width=.25\textwidth]{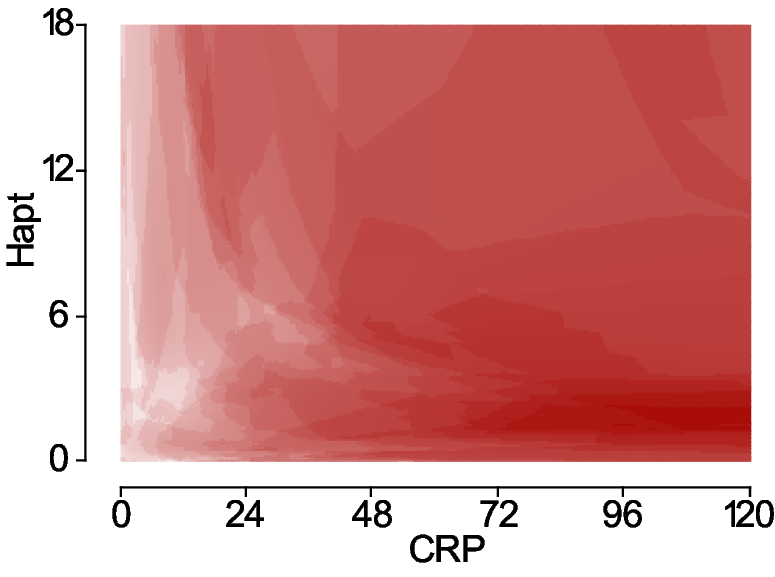}
\caption{\label{fig4}The maps of lymphoma risk for male and female dogs: (a) PDFE map for male, (b) decision tree map for male,
(c) KNN map for male, (e) PDFE map for female, (f) decision tree map for female, (g) KNN map for female, and (d) is the legend.
\textit{Disclaimer: these colored maps are for qualitative illustration and understanding and not for diagnosis
of individual patients where the more detailed maps and exact numerical values are needed.}}
\end{figure}

We use risk maps to generate hypotheses about the impact of input features. For example,
let us consider the risk of lymphoma in relation to sex for clinically suspected cohort.
There are 24 records with lymphoma and 54 records without lymphoma among female records
and 38 and 43 records with and without lymphoma correspondingly among male records in the
database of clinically suspected cases. The frequencies of lymphoma for female and male
are here 31\% and 47\% correspondingly. This probability difference can be uniformly
distributed in the space of the input attributes but can be condensed in some area on the
map. To check this hypothesis we form the risk map for the three best classifiers one of
each type for three input attributes: CHS (CRP, Hapt and sex). The best PDFE parameters
are concentrations of CRP and Hapt, 9 nearest neighbors and Gaussian kernel
(Fig.~\ref{fig4}a, e). The best decision tree parameters are linear combinations of CRP
and Hapt after logarithmic transformation, Information gain as a splitting criterion and
the weight of class with lymphoma equals 1.8 (Fig.~\ref{fig4}b, f). The best KNN options
are logarithmic transformed CRP and Hapt, Euclidean distance, 15 nearest neighbors and
Gaussian kernel for voting (Fig.~\ref{fig4}c, g).

Fig.~\ref{fig4} shows that for each classifier there are two regions: the big sex
independent area in right side and small sex dependent area in left side. In this area,
the risk of lymphoma may depend on the steroid hormones. This hypothesis needs additional
verification.

\textit{Applying the selected methods to lymphoma treatment monitoring.}
Prognosis and prediction tools give the possibility to more individualised treatments of
cancer patients \cite{ComSysBioCancer2012}. We have applied the tools we developed to the
problem of monitoring of dogs after treatment for lymphoma. The canine lymphoma blood
test was subjected to a blind retrospective study on serum collected from 57 dogs over
four years. The cLBT ranks the remission status from 0 to 5 according to PDFE lymphoma
risk evaluation, where 0 indicates complete remission, 5 equates to active diseases and a
score of 3 represents a border line result. The study demonstrated that dogs regularly
giving a cLBT score of 2 or lower remained in remission, whereas an increase in the score
to 3 or more indicated that the disease was recurring.

\begin{figure}[t]
\centerline{
\includegraphics[width=.25\textwidth]{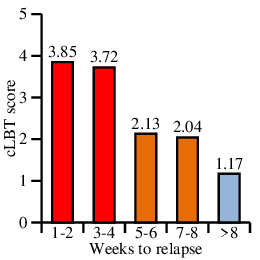}}
\caption{\label{fig5}Monitoring results: the number of weeks before relapse of lymphoma in dependence of cLBT score.}
\end{figure}

The first important result is that the score of the test immediately after treatment is
very informative for predicting the time before relapse. Fig.~\ref{fig5}a shows that for
dogs with cLBT score between 3 and 4 the time of lymphoma relapsing is about four weeks;
for dogs with cLBT score 2 the time of lymphoma relapsing is greater than four weeks and
less than eight weeks, and for dogs with cLBT score 1 the time of lymphoma relapsing is
greater than eight weeks.

The second important result is that the cLBT score indicates the relapse of lymphoma
before the clinical symptoms reappear. The study found that the test detected recurrence
up to two months prior to the appearance of physical signs. These results strongly
support the monthly basis monitoring of lymphoma patients in remission. The properly
predicted time before relapse gives the possibility for better treatment planning and we
expect that it may increase survival rate.

\section{Conclusion}\label{conclusion}
We formulate and analyze the problem of differential diagnosis of clinically suspected
cases and the problem of screening. The criteria to select the best classifier for each
problem are chosen. These criteria allow the selection of the best algorithms. For
differential diagnosis the best solution is the decision tree with three input features:
concentrations of CRP and Hapt, and Lymphadenopathy. The tree is formed with DKM as a
splitting criterion. In this tree at each node the linear combination of CRP and Hapt are
used (Fisher's approach). Synthesis of decision trees and linear discriminant analysis is
proven to be optimal in some cases. The sensitivity of the best decision tree is 83.5\%,
the specificity is 77\%.

The best result is obtained for screening by the decision tree which uses three input
features: the concentrations of CRP and Hapt, and Lymphadenopathy. CRP and Hapt are used
separately. DKM is used as the splitting criterion. The sensitivity of this method is
81.4\%, the specificity is $>$99\% (no false negative results in one-leave-out
cross-validation).

For screening on the base of two biomarker concentrations only, without any clinical
symptoms, the best decision tree uses the concentrations of CRP and Hapt separately and
Gini gain as splitting criteria. The sensitivity of this tree is 69\%, the specificity is
83.5\%.

We compare our results with some current human cancer screening tests. The accuracy of
tests which based on single biomarkers is often worse. For example, the male PSA test
gives sensitivity approximately 85\% and specificity 35\% and the CA-125 screen for human
ovarian cancer provides sensitivity approximately 53\% and specificity 98\%.
Supplementation of CA-125 by several other biomarkers increases sensitivity of at least
75\% for early stage disease and specificity of 99.7\% \cite{Rosen2005}. For the PSA
marker, using age-specific reference ranges improved the test specificity and
sensitivity, but did not improve the overall accuracy of PSA testing
\cite{HoffmanKey2002}.

The risk map visualisation is a friendly tool for explanatory data analysis. It provides
the opportunity to generate hypotheses about the impact of input features on the final
diagnosis. The risk of lymphoma (cLBT score) defined after lymphoma treatment allows prediction of
time before relapse of lymphoma. If after treatment of lymphoma the cLBT is performed
regularly, it detected recurrence up to two months prior to the appearance of physical
signs.

Canine lymphoma can be considered as a model for human non-Hodgkin lymphoma
\cite{Marconato2013}. The new diagnostic approaches can be applied for this disease.

There are several questions and directions the future work with the biomarkers CRP and Hapt for canine lymphoma
\begin{itemize}
\item{Further clinical testing of the screening classifier with special attention to the instances with lymphoma but without obvious lymphadenopathy;}
\item{Further clinical testing of the proposed lymphoma treatment monitoring system
    to validate the hypothesis that properly predicted time before relapse improves
    treatment planning and increases survival rate;}
\item{Clustering of breeds for numerosity reduction and inclusion of this important feature in the diagnostic system;}
\item{Selection of the optimal set of input features for lymphoma diagnosis from combinations of CRP and Hapt with the results of routine blood tests.}
\end{itemize}

\section{Appendix: Three main groups of algorithms}

\subsection{Decision tree}

Decision tree is a method that constructs a tree like structure which can be
used to choose between several courses of action. Binary decision trees are used in this
study. The decision tree is comprised of nodes and leaves. Every node can have a child
node. If a node has no child node it is called a leaf or a terminal node. Any decision
tree contains one root node which has no parent node. Each non terminal node calculates
its own Boolean expression (i.e. true or false). According to the result of this
calculation the decision for a given sample would be delegated to the left child node
(``true'') or to the right child node (``false''). Each leaf (terminal node) has a label
which shows how many samples of the training set belong to each class: $n_{\rm L}$ is the
number of cases with lymphoma, $n_{\rm SCWL}$ is the number of clinically suspected cases
without lymphoma, $n_{\rm H}$ is the number of healthy cases. The probability of lymphoma
is evaluated as a result of the division of the sum of weights of positive samples in
this leaf by the sum of weights of all samples in the same leaf: $$p_{\rm L}=n_{\rm L}
W_{\rm L}/(n_{\rm L} W_{\rm L}+n_{\rm SCWL} W_{\rm SCWL}+n_{\rm H} W_{\rm H}).$$ For the
screening problem $W_{\rm L}=w_{\rm L},W_{\rm SCWL}=w_{\rm L}$ and $W_{\rm H}=w_{\rm H}$.
For the problem of differential diagnosis $W_{\rm L}=w_{\rm p},W_{\rm SCWL}=1$ and
$W_{\rm H}=0$.

There are many methods to be used to develop a decision tree \cite{{RokachMaimon2010},
{Quinlan1987},{Kearne1999},{Breiman1984},{Gelfand1991},{Dietterich1996}}. We use the
methods based on information gain, Gini gain, and DKM gain. Since the screening problem
defines the prior weights of classes, these weights must be considered. There are two
ways to implement prior weights. The simplest way is to multiply the number of positive
class cases in a leaf by the weight of the positive class, and the number of negative
class cases by the weight of the negative class and then calculate the probability. In
this study we use a different method: we modify the split criteria. Let us consider one
node and one binary input attribute with values 0 and 1. To form a tree we select the
base function for information criterion among
\[
Entropy(n_{\rm L},n_{\rm n} )=-\frac{n_{\rm L}}{n_{\rm L}+n_{\rm n}}\log_2\frac{n_{\rm L}}{n_{\rm L}+n_{\rm n}}-
\frac{n_{\rm n}}{n_{\rm L}+n_{\rm n}}\log_2\frac{n_{\rm n}}{n_{\rm L}+n_{\rm n}},
\]
\[
Gini(n_{\rm L},n_{\rm n})=1-\frac{n_{\rm L}^2+n_{\rm n}^2}{(n_{\rm L}+n_{\rm n})^2}, \;
DKM(n_{\rm L},n_{\rm n})=2\sqrt{\frac{n_{\rm L} n_{\rm n}}{ (n_{\rm L}+n_{\rm n} )^2}},
\]
where $n_{\rm L}$ is the number of positive cases and $n_{\rm n}$ is the number
of negative cases. The value of the criterion is the gain of the base function:
\[
BG=Base(n_{\rm L},n_{\rm n} )-\frac{p_0+n_0}{n_{\rm L}+n_{\rm n}}Base(p_0,n_0)-
\frac{p_1+n_1}{n_{\rm L}+n_{\rm n}} Base(p_1,n_1),
\]
where $p_a$ is the number of positive cases with value of input attribute $a$,
$n_a$ is the number of negative cases with value of input attribute $a$, $Base(m,n)$
is one of the base function listed above. If each case has the weight the criterion is defined as
\[
BGW=Base(w,v)-\frac{w_0+v_0}{w+v}Base(w_0,v_0)-\frac{w_1+v_1}{w+v}Base(w_1,v_1),
\]
where $w$ is the sum of weights of positive cases, $v$ is the sum of weights of negative cases,
$w_a$ is the sum of weights of positive cases with value of input attribute equals $a$,
$v_a$ is the sum of weights of negative cases with value of input attribute equals $a$.
In this study we use $IGW$ instead of information gain, $GGW$ instead of Gini gain and $DKMW$ instead of DKM gain.

For the screening problem $w_a=w_L p_a$ and $v_a=w_{\rm L} n_{{\rm CSWL},a}+w_{\rm H}
n_{{\rm H},a}$, where $n_{{\rm CSWL},a}$ is the number of clinically suspected cases
without lymphoma with value of input attribute $a$, and $n_{{\rm H},a}$ is the number of
healthy cases with value of input attribute $a$. For the problem of differential
diagnosis $w_a=p_a, v_a=n_a$.

There are several approaches for using real valued feature for forming decision tree. The
most commonly used approach suggests the binning of the real valued attribute before form
the tree. In this study we implement the method of on the fly binning: in each node for
each real valued attribute the best threshold is searched and then this threshold is used
to bin these feature in this node. The best threshold depends on the  split criteria used
(information gain, Gini gain or DKM gain). We also use Fisher's discriminant to define
the best linear combinations of real valued features \cite{Fisher1936} in each node. This
means that we use either each real valued attribute separately or one synthetic real
valued feature instead of all real valued input attributes. Pruning techniques are
applied to improve the tree. The specified minimal number of instances in the tree's leaf
is used as a criterion to stop node splitting. This means that each leaf of the tree
cannot contain fewer instances than a specified number. For the case study we test the
decision trees which differ by:

\begin{itemize}
\item {One of the three modified split criteria (information gain, Gini gain or DKM gain);}
\item {The use of real-valued features in the splitting criteria separately or in linear combination;}
\item {The use of concentrations of Hapt and CRP or of logarithm of concentrations;}
\item {The set of input features: CH, CHA, CHL, CHS, CHAL, CHAS, CHLS and CHALS;}
\item {The minimal number of instances in each leaf is varied between 3 and 30.}
\end{itemize}

\subsection{K nearest neighbors}

The basic concept of KNN is: the class of an object is the
class of a majority of its k nearest neighbors \cite{Clarkson2005}. This algorithm is
very sensitive to distance calculation. There are several commonly used variants of
distance for KNN: Euclidean distance; Minkovsky distance; distance calculated after some
transformation of input space.

In this study we use three distances: the Euclidian distance, the Fisher's transformed
distance and adaptive distance \cite{HastieTib1996}. Moreover we use a weighted vote
procedure with weighting of neighbors by one of the standard kernel functions
\cite{Li2007}. The KNN algorithm is well known \cite{Clarkson2005}. The adaptive distance
transformations algorithm is described in \cite{HastieTib1996}. KNN with Fisher's
transformed distance is less well-known. For these methods the following options are
defined: $k$ is the number of nearest neighbors, $K$ is the kernel function, $kf$ is the
number of neighbors which are used for distance transformation. To define the risk of
lymphoma we have to do the following steps:
\begin{enumerate}
  \item Find the $kf$ nearest neighbors of test point.
  \item Calculate the covariance matrix of $kf$ neighbors and Fisher's discriminant
      direction.
  \item Find the $k$ nearest neighbors of the test point using the distance along
      Fisher's discriminant direction among the  earlier found $kf$ neighbors.
  \item Define the maximum of distances from the test point to $k$ neighbors.
  \item For each class we calculate the membership of this class as a sum of points'
      weights. The weight of a point is the product of value of the kernel function $K$ of
      distance from this point to the test point divided by maximum distance and
      predefined point weight.
  \item Lymphoma risk is defined as a ratio of the positive class membership to the sum of memberships of all classes.
\end{enumerate}
For the differential diagnosis problem the predefined weight of the lymphoma cases is equal to $w_{\rm p}$
and the predefined weight of the cases without lymphoma is equal to 1. For the screening problem the
predefined weight of clinically suspected cases is equal to $w_{\rm L}$ and for healthy cases the predefined weight is equal to $w_{\rm H}$.
The adaptive distance version implements the same algorithm but uses the other transformation on
Step 2 and other distance on Step 3. The Euclidean distance version simply defines $kf=k$ and omits  Steps 2 and 3 of the algorithm.
We test the KNN versions which differ by:
\begin{itemize}
  \item The number of nearest neighbors is varied between 1 and 20;
  \item The use of concentrations of Hapt and CRP or of logarithm of concentrations;
  \item The set of input features: CH, CHA, CHL, CHS, CHAL, CHAS, CHLS and CHALS;
  \item One of the three distances: Euclidean distance, adaptive distance and
      Fisher's distance.
  \item The kernel function for adaptive distance transformation;
  \item The kernel function for voting.
\end{itemize}

\subsection{Probability density function estimation}

We implement the radial-basis functions
method \cite{Breiman1984} for probability density function estimation \cite{Scott1992}.
For the robustness we also implement the local Mahalanobis distance transformation
\cite{Mahalanobis1936}. There are three probabilities for the screening problem:
Probability of lymphoma; Probability of belonging to the clinically suspected cohort
without lymphoma; Probability of being healthy. Each probability density function is
estimated separately by using nonparametric techniques. The total probability of lymphoma
has to be equal to the prior probability of lymphoma $p_{\rm L}^s=p$. The total
probability of belonging to the clinically suspected cohort without lymphoma is defined
by evaluation of the probability of lymphoma in the clinically suspected cohort from
data, and from the given total probability of lymphoma in population: $$p_{\rm
CSWL}^s=p_{\rm L}^s (N_{\rm CS}-N_{\rm L} )/N_{\rm L}.$$ The total probability of being
healthy is equal to 1 minus the probability of belonging to the clinically suspected
cohort: $$p_{\rm H}^s=1-p_{\rm L}^s-p_{\rm CSWL}^s.$$ For the differential diagnosis we
need to estimate two probabilities: probability of lymphoma and probability that there is
no lymphoma. The prior probabilities of these classes are defined by number of instances
in each class: $$p_{\rm L}^d=N_{\rm L}/N_{\rm CS}\mbox{  and  } p_{\rm H}^d=(N_{\rm
CS}-N_{\rm L})/N_{\rm CS}. $$

For each point, $k$ nearest neighbors from the database are defined. These $k$ points are
used to estimate the covariance matrix and calculate the Mahalanobis distance matrix.
Then the radius of the neighborhood is estimated as a maximum of the Mahalanobis
distances from data point to each of $k$ neighbors. The centre of one of the kernel
functions is placed at the data point \cite{Li2007}. The integral of any kernel function
over the whole space is equal to 1. There are $N_{\rm L}$ cases of lymphoma and $N_{\rm
L}$ kernel functions are placed at these points. The total probability is the integral of
the sum of kernel functions and is equal to $N_{\rm L}$ but the total probability of
lymphoma has to be equal to the prior probability $p_{\rm L}^t$ (where $t$ is `s' for the
screening problem and `d' for differential diagnosis problem). It means that the sum of
kernel functions has to be multiplied by $W_{\rm L}=p_{\rm L}^t/N_{\rm L}$.

The probability of lymphoma at an arbitrary point is estimated as products of weight $W_{\rm L}$
and the sum of values of kernel functions which are placed at data points that
correspond to records with lymphoma. Other probabilities are estimated analogously.

We use the following steps to evaluate the risk of lymphoma: (i) three (screening problem) or
 two (differential diagnosis problem) probabilities are estimated and (ii) the risk of lymphoma is
 defined as a ratio of the probability of lymphoma to the sum of all probabilities.
We test the PDFE versions which differ by:
\begin{itemize}
  \item The number of nearest neighbors (it is varied between 5 and 30);
  \item The use of concentrations of CRP and Hapt or logarithm of the concentrations;
  \item The set of input features: CH, CHA, CHL, CHS, CHAL, CHAS, CHLS or CHALS;
  \item The kernel function which is placed at each data points.
\end{itemize}

\subsection{Computational cost}

Let us compare the computational cost of the most expensive procedure, LOOCV, for these
three types of algorithms. All software has been implemented in Java 6 with one core
usage. A computer with processor Intel(R) Core(TM) i7-3667U CPU 2.0GHz 2.5GHz with 8GB RAM
under 64-bit Windows 7 Enerprise operation system has been used. The test results are
presented in Table~\ref{tab5}. This is the time for LOOCV of one model. For selection
of the best decision tree this LOOCV routine was called 10,368 times for the screening problem and
5,184,400 times for the differential diagnosis, for the best KNN method it was called
25,600,000 times and 3,840 times for the best PDFE.

\begin{table}{\footnotesize\begin{center}
\caption{\label{tab5} LOOCV time for one model}
\begin{tabular}{cc}
\hline
Classifier& Time (sec)  \\   \hline
Decision tree & 0.22  \\  \hline
KNN & 0.00005 \\   \hline
PDFE & 0.14 \\
 \hline
\end{tabular}
\end{center}}
\end{table}

\newpage

{\bf Summary}

Lymphoma is one of the most frequent canine cancers. It can be also considered as a model
for human non-Hodgkin lymphoma. We develop technology for differential diagnosis of
canine lymphoma, for screening and for remission monitoring. This technology is based on
a specific blood test.

The canine lymphoma blood test detects the levels of two biomarkers, the acute phase
proteins, C-Reactive Protein and Haptoglobin.  This test can be used for diagnostics, for
screening, and for remission monitoring. We analyze clinical data, test various machine
learning methods and select the best approach to these problems.

Three family of methods, decision trees, kNN (including advanced and adaptive kNN) and
probability density evaluation with radial basis functions, are used for classification
and risk estimation. Several pre-processing approaches were implemented and compared. The
best of them are used to create the diagnostic system. For the differential diagnosis the
best solution gives the LOOCV sensitivity and specificity of 83.5\% and 77\%,
respectively (using three input features, CRP, Haptoglobin and the standard clinical
symptom). For the screening task, the decision tree method provides the best result, with
sensitivity and specificity of 81.4\% and $>$99\%, respectively (using the same input
features), and if the clinical symptoms (Lymphadenopathy) are considered as unknown then
a decision tree with CRP and Hapt provides sensitivity 69\% and specificity 83.5\%.

The lymphoma risk evaluation problem is formulated and solved. We use three methods to
evaluate risk. The best models are selected as the system for computational lymphoma
diagnosis and evaluation the risk of lymphoma as well. These methods are implemented into
a special web-accessed software and are applied to problem of monitoring dogs with
lymphoma after treatment. It detects recurrence of lymphoma up to two months prior to the
appearance of clinical signs and may help to optimize relapse treatment. The risk map
visualisation provides a friendly tool for explanatory data analysis.

We compare our results with some current human cancer screening tests. The accuracy of
tests which based on single biomarkers is often worse. For example, the male PSA test
gives sensitivity approximately 85\% and specificity 35\% and the CA-125 screen for human
ovarian cancer provides sensitivity approximately 53\% and specificity 98\%.
Supplementation of the tests by several other biomarkers increases sensitivity and
specificity.

\end{document}